\documentstyle[aps,twocolumn,epsfig]{revtex}
\newcommand \bea{\begin{eqnarray}}
\newcommand \eea{\end{eqnarray}}

\newcommand \la{\raisebox{-.5ex}{$\stackrel{<}{\sim}$}}
\newcommand{\av}[1]{\langle{#1}\rangle}

\begin{document}
\title{Anomalous Multiplicity Fluctuations from Phase Transitions in
Heavy Ion Collisions}
\author{H. Heiselberg$^{a}$ and A. D. Jackson$^{b}$}
\address{$^{a}$NORDITA, Blegdamsvej 17, DK-2100 Copenhagen \O, Denmark\\
$^{b}$NBI, Blegdamsvej 17, DK-2100 Copenhagen \O, Denmark}

\maketitle

\begin{abstract}
Event-by-event fluctuations and correlations between particles
produced in relativistic nuclear collisions are studied.  The
fluctuations in positive, negative, total and net charge are closely
related through correlations.  In the event of a phase transitions to
a quark-gluon plasma, fluctuations in total and net charge can be
enhanced and reduced respectively which, however, is very sensitive to
the acceptance and centrality. If the colliding system experiences
strong density fluctuations due, e.g., to droplet formation in a
first-order phase transition, all fluctuations can be enhanced
substantially.  The importance of fluctuations and correlations is
exemplified by event-by-event measurement of the multiplicities of
$J/\Psi$'s and charged particles since these observables should
anti-correlate in the presence of co-mover or anomalous
absorption. \\[10mm]
\end{abstract}




\vspace{1cm}

Event-by-event fluctuations have been measured at the SPS \cite{NA49,WA98} 
and, with the higher multiplicities of RHIC and LHC, will become 
an important 
tool for studying the anomalous fluctuations and correlations that might 
remain following the phase transition to a quark-gluon plasma (QGP).  Studies 
of event-by-event fluctuations at SPS energies do not indicate the presence 
of new physics \cite{SRS,BH,JK}. It has, however, been proposed that large
multiplicity fluctuations can arise from density fluctuations or
droplets created by a first-order phase transition \cite{BH}.  Recently, 
it has also been suggested that fluctuations in net charge should 
be suppressed in a QGP \cite{JK2,AHM,Hphysrep}. 
Here, we shall consider multiplicity fluctuations in some generality 
to see how they are affected by conservation of total charge and 
strangeness, to understand the correlations between various measured 
fluctuations, and to show how their measurement can reveal interesting 
details of the collision process.

{\bf Multiplicity fluctuations} between various kinds of particles can 
be strongly correlated. As an example, 
consider the multiplicities of positive and negative pions, $N_+$ and $N_-$, 
in a rapidity interval $\Delta y$ for any relativistic heavy-ion experiment.  
(Similar analyses can be performed for any two kinds of 
distinguishable particles.)  We define the fluctuation in any multiplicity 
$N$ as
\bea
   \omega_N\equiv \frac{\av{N^2}-\av{N}^2}{\av{N}} \,. \label{o}
\eea
Empirically, the fluctuations of Eq.\,(\ref{o}) are
typically of order unity in relativistic nuclear collisions which
is consistent with the expectations of Poisson statistics.

The net positive 
charge from the protons in the colliding nuclei is much smaller than the 
total charge produced in an ultrarelativistic heavy-ion
collision.  For example, $\av{N_+}$ exceeds $\av{N_-}$ by 
only $\sim15$\% at in Pb$+$Pb collisions at SPS energies.  The fluctuations 
in the number of positive and negative (or neutral) pions are 
also very similar, $\omega_{N_+}\simeq\omega_{N_-}$.
Charged particle fluctuations have been estimated in thermal as well
as participant nucleon models \cite{BH} including effects of
resonances, acceptance, and impact parameter fluctuations.  By varying
the acceptance and centrality, the degree of thermalization can
actually be determined empirically \cite{Hphysrep}. Detailed analysis
indicates that the fluctuations in central Pb+Pb 
collisions at the SPS are thermal whereas peripheral collisions are a 
superposition of pp fluctuations \cite{NA49v}.

The fluctuations in the total ($N_{ch}=N_+ + N_-$) and net ($Q=N_+-N_-$) 
charge are defined as
\bea
 && \frac{\av{(N_+\pm N_-)^2}-\av{N_+\pm N_-}^2}{\av{N_++ N_-}}
 = \nonumber\\
 &&\quad \frac{\av{N_+}}{\av{N_{ch}}}\omega_{N_+}
    +\frac{\av{N_-}}{\av{N_{ch}}}\omega_{N_-} \, \pm \, C \,, \label{oNpm}
\eea
where the correlation is given by
\bea
  C = \frac{\av{N_+N_-}-\av{N_+}\av{N_-}}{\av{N_{ch}}/2} \,. \label{C}
\eea
Fluctuations in positive, negative,
total and net charge can be combined to yield 
both the intrinsic fluctuations in the numbers of $N_{\pm}$ and the 
correlations in their production as well as a consistency check. 
These quantities can change as a 
consequence of thermalization and a possible phase transition.

In practice, $\omega_{N_+}\approx \omega_{N_-}$, 
so that the fluctuation in total charge simplifies to
\bea
  \omega_{N_{ch}} &\equiv& \frac{\av{N_{ch}^2}-\av{N_{ch}}^2}{\av{N_{ch}}}
   =\omega_{N_+} + C \,, \label{oNch}
\eea
and that for the net charge becomes 
\bea
  \omega_Q &\equiv& \frac{\av{Q^2}-\av{Q}^2}{\av{N_{ch}}}
   =\omega_{N_+} - C \,. \label{oQ}
\eea

The fluctuation in net charge is related to the fluctuation in the
ratio of positive to negative particles,
$\omega_Q\simeq\av{N_+/N_-}\av{N_{ch}}\omega_{N_-/N_+}/4$ plus volume
(or impact parameter) fluctuations \cite{JK,JK2}.  The virtue of this
expression is that volume fluctuations can in principle be extracted
empirically. Alternatively one can vary the centrality bin size or
the acceptance \cite{Hphysrep}. In the following we shall assume that
such ``trivial'' volume fluctuations have been removed.

The analysis has so far been general and Eqs. (\ref{oNpm}-\ref{oQ})
apply to any kind of distinguishable particles, e.g. positive and
negative particles, pions, kaons, baryons, etc. - irrespective of what
phase the system may be in, or whether it is thermal or not.  In the
following, we shall consider thermal equilibrium, which seems to apply
to central collisions between relativistic nuclei, in order to reveal
possible effects on fluctuations of the presence of a quark-gluon
plasma.

Bosons/fermions have thermal fluctuations, $\omega_N=1\pm\av{n_p^2}/\av{n_p}$
where $n_p=(\exp(\epsilon_p/T)\mp1)^{-1}$ is the boson/fermion distribution
function, which are slightly larger/smaller than those
of Poisson statistics for a Boltzmann 
distribution. Massless bosons, e.g.\ gluons, have 
$\omega_B=\xi(2)/\xi(3)=1.37$ and massless fermions, 
e.g.\ quarks, have $\omega_F=2\xi(2)/3\xi(3)\simeq0.91$ 
independent of temperature.  
Massive bosons have smaller fluctuations with, for example, $\omega_\pi=1.11$ 
and $\omega_\rho=1.01$ when $T=m_\pi$.

{\bf In a thermal hadron gas} (HG) as created in relativistic 
in nuclear collisions, pions can be produced either directly or through the 
decay of heavier resonances, $\rho,\ \omega,...$. The resulting
fluctuation in the measured number of pions is
\bea
  \omega_{N_+}=\omega_{N_-} = f_\pi\omega_\pi +f_\rho\omega_\rho+
   f_\omega\omega_\omega + .... \label{oN} \, ,
\eea
where $f_r$ is the fraction of measured pions produced 
from the decay of resonance $r$, and $\sum_r f_r=1$. 
These mechanisms are 
assumed to be independent, which is valid in a thermal system. 

The heavier resonances such as $\rho^0, \omega,...$
decay into pairs of $\pi^+\pi^-$ and thus lead to a correlation
\bea
 C^{HG} = \frac{1}{3}f_\rho +f_\omega + .... \,. \label{CHG}
\eea
Resonances reduce the fluctuations in net charge in a
HG to $\omega_Q=0.70$ \cite{JK,SRS}.
In addition, total charge neutrality reduces fluctuations in net charge
when the acceptance is large and thus increases
correlations as will be discussed below.

{\bf A phase transition to the QGP}
can alter both fluctuations and correlations in the 
production of charged pions.  To the extent that these effects are not 
eliminated by subsequent thermalization of the HG, they may remain as 
observable remnants of the QGP phase.  
As shown in Refs.\,\cite{JK2,AHM}, net charge fluctuations in a plasma 
of {\it u,\ d} quarks and gluons are reduced partly due to the intrinsically 
smaller quark charge and partly due to correlations from gluons
\bea
 \omega_Q  = \frac{\av{N_q}}{\av{N_{ch}}} \omega_F  
\frac{1}{N_f} \sum_{f=u,d,...}^{N_f} q_f^2 
  \,, \label{oq}
\eea
where $N_f$ is the number of quark flavors, $q_f$ their charges, and
$N_q$ the number of quarks.
The total number of charged particles (but not the net charge) can 
be altered by the ultimate hadronization of the QGP.  This effect can 
be estimated by equating the entropy of all pions to the entropy of the 
quarks and gluons.  
Since 2/3 of all pions are charged and since the entropy 
per fermion is 7/6 times the entropy per boson in a QGP
\bea
  \av{N_{ch}} \simeq \frac{2}{3}(\av{N_g} + \frac{7}{6}\av{N_q}) \, ,
 \label{oq2}
\eea
where the number of gluons is $\av{N_g}=(16/9N_f)\av{N_q}$.
Inserting this result in (\ref{oq}), we see that the resulting fluctuations 
are $\omega_Q=0.18$ in a two-flavor QGP 
(and $\omega_Q=0.12$ for three flavors). 
As pointed out in \cite{JK2}, lattice results give 
$\omega_Q \simeq 0.25$. 
However, according to \cite{SH} a substantial fraction of the
pions are decay products from the HG, and the
entropy of the HG exceed that of a pion gas by a
factor $1.75-1.8$. As described in \cite{AHM}
the net charge fluctuations
should be increased by this factor in the QGP, i.e.
$\omega_Q\simeq0.33$ in a two-flavor QGP, whereas it remains similar in the
HG, $\omega_Q\simeq0.6$.

However, if the high density phase is initially dominated by gluons with 
quarks produced only at a later stage of the expansion by gluon fusion, the 
production of positively and negatively charged quarks will be strongly 
correlated on sufficiently small rapidity scales.  If, for example, 
the entropy density increases by an order of magnitude in going from a HG 
to QGP without additional net charge production, fluctuations 
in net charge will be reduced significantly,
\bea
   \omega_Q^{QGP}\simeq \frac{s_{HG}}{s_{QGP}} \, \omega_Q^{HG}. \label{os}
\eea
The resulting fluctuation in net charge is necessarily {\it smaller}
than that from thermal quark production as given by Eq.\,(\ref{oq}).
A similar phenomenon occurs in string models where particle production
by string breaking and $q\bar{q}$ pair production results in flavor
and charge correlations on a small rapidity scale \cite{BF,Whitmore}.  The
recently measured charged particle density at midrapidity in central
nuclear collisions at RHIC \cite{PHOBOS} is only $\sim30-40$\% larger
than $pp$ scaled up by the nuclear mass as was also found at SPS
energies \cite{NA49}.  If entropy and multiplicities are proportional,
the net and total charged particle fluctuations should be the same as
at SPS according to Eq. (\ref{os}) unless anomalous non-thermal
fluctuations occur as will be discussed below.

The {\it strangeness} fluctuation in kaons $K^\pm$ might seem less
interesting at first sight since strangeness is not suppressed in the
QGP: The strangeness per kaon is unity, and the total number of kaons
is equal to the number of strange quarks.  However, if strange quarks
are produced at a late stage in the expansion of a fluid initially
dominated by gluons, the net strangeness will again be greatly reduced
on sufficiently small rapidity scale.  Consequently, fluctuations in
net/total strangeness would be reduced/enhanced.

The {\it baryon number} fluctuations have been estimated in a thermal
model \cite{AHM}. It is, however, not known how the annihilation of
baryons and antibaryons in the hadronic phase affect these results.
If only charged particles are detected, but not 
$K^0$, $\bar{K}^0$, neutrons and antineutrons,
the fluctuations have smaller correlations 
as compared to the total and net strangeness or baryon number.

{\bf Total charge conservation} is important when the acceptance
$\Delta y$ is a non-negligible fraction of the total rapidity.  It
reduces the fluctuations in the net charge as calculated within the
canonical ensemble, Eqs.\,(\ref{oN},\ref{CHG}-\ref{oq2}).  If the total
positive charge (which is exactly equal to the total negative charge plus the
incoming nuclear charges) is
independently distributed according to the single particle distributions, 
the resulting fluctuations within the acceptance are 
\bea
   \omega_Q &=& 1-f_{acc} \,, \label{oQrandom} \\
   \omega_{N_{ch}} &=& 1-f_{acc}+2f_{acc}\,\omega_{N_+} \, \label{orandom}
\eea
where $f_{acc}=N_{tot}^{-1}\int_{\Delta y}(dN_{ch}/dy)dy$ is the
acceptance fraction or the probability
that a charged particle falls into the acceptance $\Delta y$.
Since charged particle rapidity distributions
are peaked near midrapidity, charge conservation effectively kills
fluctuations in the net charge even when $\Delta y$ is substantially
smaller than the laboratory rapidity, $y_{lab} \simeq 6$ (11) 
at SPS (RHIC) energies. Total charge conservation also has the effect of 
increasing $\omega_{ch}$ towards $2\omega_{N_+}$ according to 
Eqs.\,(\ref{oNch}) and (\ref{orandom}). Similar effects can be 
seen in photon fluctuations when photons are produced in
pairs through $\rho^0 \to 2\gamma$. In the WA98 experiment,
$\omega_\gamma\simeq 2$ is found after acceptance corrections
and eliminating volume fluctuations \cite{WA98}.

When the acceptance $\Delta y$ is too small, particles from
a thermal ensemble can 
diffuse in and out of the acceptance
during hadronization and freezeout \cite{AHM}.  Furthermore, 
correlations due to resonance production will disappear when the average
separation in rapidity between decay products exceeds the acceptance.
Each of these effects tend towards Poisson
statistics when $\Delta y\la\delta y$, where $\delta y$ 
denotes the rapidity interval that particles diffuse during hadronization,
freezeout and decay. 
If $\omega_Q$ is the canonical thermal fluctuation of 
Eqs.\,(\ref{CHG},\ref{oq2}), the resulting fluctuation
after correcting for both $\delta y$ and total charge conservation is
approximately
\bea
   \omega_Q^{exp} \simeq \left(\frac{\Delta y}{\Delta y+2\delta y}\omega_Q
  + \frac{2\delta y}{\Delta y+2\delta y} \right) (1-f_{acc}) 
  \,.  \label{oQexp}
\eea
Here, the factor $(1-f_{acc})$ is due to total charge conservation
as in (\ref{oQrandom}). The remainder is fluctuations from
two sources: a thermal one with fluctuations $\omega_Q$ and a random
one with Poisson fluctuations, each weighted with the fraction of
the charged particles they contribute.

The resulting fluctuations in total and net charge are shown in
Fig.\,1 assuming $\omega_{N_+}=\omega_\pi\simeq1.1$ and $\delta
y=0.5$. As mentioned above, $f_{acc}$ and $\Delta y$ are related by
the measured charge particle rapidity distributions
\cite{NA49}. Preliminary NA49 data \cite{NA49,JK} agree well with the
net and total fluctuations in a HG ($C=0.4$) from Eqs.\,(\ref{oQexp})
and \,(\ref{oq2}).  Residual volume fluctuations are significant for
$\omega_{N_{ch}}$ \cite{BH} and have been estimated and subtracted.
The curves apply to RHIC energies as well after scaling $\delta y$
with $\Delta y$.

\vspace{-1cm}
\begin{figure}
\centerline{\psfig{figure=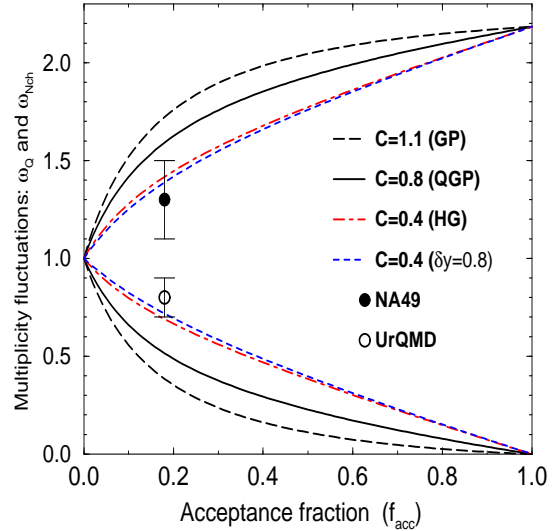,width=8cm,height=8cm,angle=0}}
\vspace{1cm} 
\caption{Acceptance dependence of thermal fluctuations in net charge 
($\omega_Q$ of Eq.\,(\ref{oQexp}), lower curves)
and  total ($\omega_{N_{ch}}$, upper curves).
Correlations increase from a hadron gas ($C\simeq 0.4$), to a QGP
($C\simeq0.8$) and a pure gluon plasma ($C\simeq1.1$) (see text). 
\label{fig1}  }
\end{figure}

The net charge fluctuations in a thermal HG corrected for finite
acceptance and diffusion are slightly below the value without any
intrinsic correlations given by Eq. (\ref{oQrandom}).  This reduction
is due to the correlations in the HG that leads to $\omega_Q<1$. In
high energy $pp$ collisions there are stronger rapidity correlations
between unlike than like charged particles \cite{BF,Whitmore} leading
to similar magnitude for $\omega_Q$
\footnote{The correlations are related to the two-body density
distributions, e.g.,
$\av{N_+N_-}=\int\rho^{(2)}_{+-}(y_1,y_2)dy_1dy_2$, where the integral
extends over $y_1,y_2\in\Delta y$.}. Therefore the net charge
fluctuations does not vary by much going from peripheral $pp$-like
high energy nuclear collisions to central collisions that are more
likely to produce a thermal hadronic gas.
The fluctuations in total charge are, however, very different
because the total
charge fluctuations in $pp$ collisions increase dramatically with
collision energy, $\omega_{N_{ch}}^{RHIC}\simeq 6$
and  $\omega_{N_{ch}}^{LHC}\simeq 20$
as compared to $\omega_{N_{ch}}^{SPS}\simeq 2.0$ \cite{BF,Whitmore}.
Peripheral collisions will therefore be very different from central
ones and the centrality dependence should be studied carefully to
assess the degree of thermalization before anomalous fluctuations due
to phase transitions can be determined \cite{Hphysrep}.

{\bf Large non-thermal fluctuations} can arise as a consequence of
density fluctuations due, e.g.\ to droplet 
formation in first-order phase transitions.  These could lead to large 
fluctuations in multiplicities \cite{BH,Hphysrep,HJ} of charged particles 
and therefore also in the total and net charge. 
The above estimates for the fluctuations were of order unity.
They implicitly assumed a uniform 
expanding system.
Consider a scenario where the total
multiplicity within the acceptance arises from a normal hadronic
background component ($N_{HG}$) and from a second component
($N_{QGP}$) that has undergone a transition:
\bea
  N = N_{HG}+N_{QGP} \,. \label{NHQ}
\eea
Its average is $\av{N}=\av{N_{HG}}+\av{N_{QGP}}$.  Assuming that
the multiplicity of each of these components is statistically 
independent, the multiplicity fluctuation becomes
\bea
  \omega_N = \omega_{HG} + (\omega_{QGP}-\omega_{HG})
       \frac{\av{N_{QGP}}}{\av{N}} \,. \label{omtr}
\eea
Here, $\omega_{HG}$ is the standard fluctuation in
hadronic matter $\omega_{HG} \simeq 1$. The fluctuations due to the 
component that had experienced a phase transition, $\omega_{QGP}$, depend 
on the type and order of the transition, the speed with which the collision 
zone goes through the transition, the degree of equilibrium, the subsequent 
hadronization process, the number of rescatterings between hadronization and
freezeout, etc.
If  thermal and chemical equilibration eliminate 
all signs of the transition, $\omega_{QGP} \simeq \omega_{HG}$.  At the 
other extreme, the droplet scenario could produce 
$\omega_{QGP}\sim\av{N}\sim 10^2-10^3$ if most hadrons arrive from a
droplet so that either all or none fall into the acceptance \cite{BH}.  This 
is a promising signal worth looking for.  Since
droplets or density fluctuations are expected to 
be charge neutral, net charge fluctuations should vanish 
$\omega_Q\simeq 0$ whereas $\omega_{ch}\simeq2\omega_{N_+}\sim2\omega_{QGP}$.

{\bf General correlators} between all particle species 
should be measured event-by-event, e.g., the ratios \cite{BH}
\bea
  \frac{\av{N_i/N_j}}{\av{N_i}/\av{N_j}} \simeq 
       1+\frac{\omega_{N_j}}{\av{N_j}} 
       - \frac{\av{N_iN_j}-\av{N_i}\av{N_j}}{\av{N_i}\av{N_j}}   \,,
   \label{ratio}
\eea
where $N_{i,j}$ are the multiplities in acceptances $i$ and $j$ of 
any particle type.
(We assume that $N_j$ is so large that it never vanishes.)
 In the presence of droplets, $N_i$ and 
$N_j$ would be strongly correlated in nearby rapidity intervals and at 
all azimuthal angles.

As another example, consider correlations between multiplicities of
charmonium particles $N_\psi$, $\psi=J/\Psi,\psi',\chi,..$ and charged
particle multiplicity ($N$) in a {\bf given} rapidity interval $\Delta
y$.  If a $\psi$ is absorbed on co-movers or anomalously suppressed by
QGP, one would expect {\it anti-correlations} because the number of
co-movers and QGP should scale with the multiplicity $N$.  By contrast,
direct Glauber absorption should not depend on the multiplicity of
particles in $\Delta y$ for a given centrality since it is the result
of collisions with participating nucleons in Glauber trajectories
along the beamline.

To quantify this anti-correlation, we model the absorption of $\psi$s
by simple Glauber theory
\bea
  \frac{\bar{N}_\psi}{N_\psi^0}= e^{-\sigma_{c\psi}\rho_c l} 
   \equiv  e^{-\gamma N/\av{N}} \,,
\eea
where $\bar{N}_\psi$ is the average number of $\psi$'s for given
charge particle multiplicity $N$, and
$N_\psi^0$ is the number of $\psi$s before co-mover or anomalous
absorption sets in but after direct Glauber absorption on participant
nucleons.  In Glauber 
theory, the exponent is the product of the absorption cross
section $(\sigma_{c\psi}$), the absorber density $(\rho_c)$, and
the average path length $(l)$ traversed in matter.  The 
density, therefore also the exponent, is proportional to the multiplicity $N$
with coefficient $\gamma=-d\log N_\psi/d\log N$. In a simple co-mover
absorption model for a system with longitudinal Bjorken 
scaling, $\gamma$ can be calculated to be approximately \cite{HM}
\bea
  \gamma\simeq \sum_c\frac{dN_c}{dy} 
\frac{\av{v_{c\psi}\sigma_{c\psi}}}{4\pi R^2}  \log(R/\tau_0) \,,
\eea
where $dN_c/dy$ is the co-mover rapidity density, $\sigma_{c\psi}$ the
absorption cross section, $v_{c\psi}$ the relative velocity, $R$ the
transverse size of the overlap zone, and $\tau_0$ the formation time.
Co-Mover absorption reduces the number of $\psi$ by a factor $e^{-\gamma}$
where $\gamma$ increases with centrality up to $\gamma \sim 1$ for typical 
parameters employed in co-mover absorption models \cite{Gavin}.

Since fluctuations in the exponent are small, $\gamma\sqrt{\omega_N/\av{N}} 
\ll 1$, the anti-correlation is
\bea
  \frac{\av{NN_\psi} -\av{N}\av{N_\psi}}{ \av{N_\psi}} = \,
 -\gamma \omega_N \,. \label{NNpsi}
\eea
It is negative and proportional to the amount of co-mover and anomalous 
absorption.  It vanishes when the absorption is independent of multiplicity
($\gamma=0$).  The rapidity interval should not be less than the typical
relative rapidities between co-movers and the $\psi$
or the rapidity range of a droplet.

Even in central heavy-ion collisions $\psi$'s are rarely produced and
so $N_\psi=1$ or 0. In both cases one should measure $dN_{ch}/dy$ and
average over events with and without a $\psi$ separately. If comovers
absorb the $\psi$ we expect that $\langle dN_{ch}/dy\rangle$ is
slightly smaller for the events with a $\psi$ than without, leading to
the negative correlation in Eq. (\ref{NNpsi}).

A quantitative assessment of the average suppression of $\psi$s due to
co-mover absorption versus direct $\psi$-nucleon absorption has been
debated ever since the first measurements of $J/\Psi$ suppression.
The anticorrelations of Eq.\,(\ref{NNpsi}) directly quantify the
amount of co-mover or anomalous absorption and can therefore be
exploited to distinguish between these and direct Glauber absorption
mechanisms.  In that respect it is similar to the elliptic flow
parameter for $\psi$ \cite{HM}. 
For a sample of $N^\psi_{events}$, 
the statistical uncertainty in $\gamma$ as determined by Eq.\,(\ref{NNpsi}) 
is $\sim \sqrt{\av{N}/N^\psi_{events}}$. If we take a rapidity bin $\Delta y 
\sim 1$ and consider central heavy ion collisions, $\av{N}$ will range 
from $\sim 10^2-10^3$ in going from SPS to RHIC energies. 
A sample of $10^4-10^5$ $\psi$s would be 
sufficient to determine $\gamma$ with an accuracy of $\pm 0.1$.
The analysis of the kaon to pion ratio by NA49 obtains such an
accuracy by comparing to a mixed event analysis which removes systematic
errors \cite{NA49}.

The impact parameter fluctuations and correlations may not cancel
exactly for the $\psi/N_{ch}$ ratio, as they do for the $\pi^+/\pi^-$
ratio \cite{JK}, because their production mechanisms differ. It is
hard for the $\psi$ and soft for most of the charged particles and
thus their multiplicities scale approximately with the number of
binary NN collisions and the number of participants respectively.  The
number of NN collisions increase more rapidly with centrality and
nuclear mass number $(\propto A^{4/3}$. As a result the impact parameter
correlations between the $\psi$ and $N$ in Eq. (\ref{NNpsi}) will be
slightly larger than the impact parameter fluctuations implicit in the
second term in Eq. (\ref{ratio}).  The difference is a finite fraction
of the total impact parameter fluctuations and will be of similar
magnitude but opposite sign as correlations from co-mover absorption.
The net impact parameter fluctuations will, however, depend
on centrality and decrease as bin-size of the centrality cut decreases
which should make it possible to separate it from other correlations.
It would reveal independent information on the soft vs. hard
production mechanisms. 

{\bf In summary,} we have given a detailed analysis 
of total and net charge fluctuations and correlations including total charge
conservation and diffusion effects and how they depend on the
acceptance.  The correlations may increase if a QGP is formed
resulting in a reduction/enhancement of net/total charge by up to an
order of magnitude depending on the model. It is important to measure
fluctuations and correlations for various acceptances as well as
versus centrality and/or beam energy. At SPS energies the fluctuations
in net charge actually {\it increase} slightly 
with centrality \cite{NA49v} due
to thermalization which is opposite to the predicted decrease in net
charge fluctuations if a QGP is formed in central heavy-ion
collisions.  
At RHIC and LHC energies the correlations vs. centrality
will be much larger because the total
charge fluctuations in $pp$ collisions are $\omega_{N_{ch}}^{RHIC}\simeq 6$
and  $\omega_{N_{ch}}^{LHC}\simeq 20$
as compared to $\omega_{N_{ch}}^{SPS}\simeq 2.0$ \cite{BF,Whitmore}.
Peripheral collisions will therefore be very different from central
ones and the centrality dependence should be studied carefully to
assess the degree of thermalization before anomalous fluctuations due
to phase transitions can be determined \cite{Hphysrep}.
It is important to measure all multiplicity fluctuations
and correlations and understand the physical effects discussed above
before a possible small increase in correlations can be attributed to
the formation of a QGP.

We are grateful for discussions with S. Voloshin (NA49) and T. Nayak
(WA98) and for showing us preliminary data.

\end{document}